\documentclass[aps,pra,showpacs,amsmath,reprint,twocolumn]{revtex4-1}

\usepackage{graphicx}
\usepackage{url}
\usepackage{wasysym}

\newcommand{\pd}{\partial}
\newcommand{\bs}{\boldsymbol}

\begin{document}


\title{Dirac Kirchhoff diffraction theory} 



\author{Ruben Van Boxem}
\email{ruben.vanboxem@ua.ac.be}
\author{Bart Partoens}
\email{bart.partoens@ua.ac.be}
\author{Jo Verbeeck}
\email{jo.verbeeck@ua.ac.be}
\affiliation{EMAT \& CMT, University of Antwerp, Groenenborgerlaan 171, 2020 Antwerp, Belgium}


\date{\today}

\begin{abstract}
Kirchhoff's scalar diffraction theory is applied throughout photon and electron optics.
It is based on the stationary electromagnetic or Schr\"odinger wave equation, and is useful in describing interference phenomena for both light and matter waves.
Here, Kirchhoff's diffraction theory is derived from the relativistic Dirac equation, thus reformulated to work on Dirac spinors.
The parallels with the ``classic" scalar theory are highlighted, and a basic interpretation of the result obtained for the Fraunhofer diffraction limit is given.
The goal of this paper is to emphasize the similarity between the two equations on the most fundamental level.
\end{abstract}


\maketitle 

\section{Introduction}

Scalar diffraction theory is what lies at the basis of Fourier optics, and is applied in many different fields in optics~\cite{Henault,Takeda,Takeda2}.
The formal mathematical theory of diffraction was developed by Gustav Robert Kirchhoff in the late 1800's.
He succeeded in removing the arbitrariness in the Huygens-Fresnel theory.
Later, Rayleigh and Somerfeld removed the mathematical inconsistency formed by Kirchhoff's boundary conditions.
This led to two distinct possible solutions, depending on what boundary conditions are chosen for the physical model.
Kirchhoff's original result was found to be the arithmetic average of the two Rayleigh-Somerfeld solutions.
For simplicity, we consider only Kirchhoff's original premise here.
\\
While investigating electron diffraction through an aperture, which recently led to the experimental observation of electron vortex beams ~\cite{Verbeeck, Verbeeck_spiral, Schattschneider_fork}, the question arose whether the simple description of the diffraction process by a single Fourier transform was also valid for relativistic fermions.
This goes along with the question of what happens to the particle's spin.
Both issues can be answered by solving this problem using the Dirac equation.
There is work being done on developing the theory behind electron vortices~\cite{Bliokh_semiclassical, Schattschneider_theory}, and there have been previous developments with respect to the holographic generation of vortex beams~\cite{Janicijevic}.
A fully relativistic description of the creation of electron vortices is useful to ascertain the exact form of the incoming waves in scattering experiments performed using a wave created by a special aperture.
Usually, scattering theory uses plane waves to describe the interaction process, but a vortex cannot possibly be simply approximated to a plane wave.
A proper form for the wave created by binary holograms is required if one is ever to develop a fully relativistic description of the electron vortex scattering process as it happens in current experiments.
\\
Explicit calculation of the optical diffraction theorems based on the Dirac equation led to a pleasing result: the Dirac equation lends itself to a generalization of the classical result, containing variables like spin through the introduction of 4-component Dirac spinors.
The formulas obtained here can be used to derive more precise calculations based on the Dirac equation for a variety of fermion wave optics experiments, only by relaxing some approximations assumed here for clarity.
The derivations in this work are based on Hillion's work on the spinor description of polarized light~\cite{Hillion_Fresnel, Hillion_Helmholtz, Hillion_geometric, Hillion_spinor}, which at its core, uses a 2-component massless Dirac equation.
\\
Below, a short derivation of the scalar diffraction theory is given containing the essential elements required to generalize it for the Dirac case.
Afterwards, the exact relation between the Kirchhoff and Rayleigh-Somerfeld theories will be discussed as it pertains to the Dirac equation.
Natural units $\hbar=c=1$ are used throughout, shorthand partial derivative notation is used: $\partial_{x^i} = \frac{\partial}{\partial x^i}$, and Einstein summation is implied over three (Latin letters) or four (Greek letters) dimensions where appropriate.
For a more rudimentary derivation of the scalar theory, refer to your favorite book on optics or Hillion's paper on the spinor Helmholtz equation~\cite{Hillion_Helmholtz}, where a 2-component spinor form of EM waves is used.

\section{Scalar diffraction theory}

\subsection{Helmholtz-Kirchhoff integration theorem}

The time-independent Helmholtz equation can be derived from the stationary Schr\"odinger equation as well as the optical wave equation for the scalar electric field $E$.
In what follows, the quantum mechanical notation of $\Psi$ is used, to ease the transition to Dirac spinors.
The static wave equation for an eigenstate $\Psi$ with energy $E=\frac{\hbar^2k^2}{2m}$ is given by:
\begin{equation} \label{eq:scalar_wave_equation}
\nabla^2 \Psi(\bs r) + k^2 \Psi(\bs r) = 0.
\end{equation}
This is the static Helmholtz equation.
Setting the right-hand side to $-\delta(\bs r - \bs r')$, and keeping the part of the solution that satisfies Sommerfeld's radiation condition~\cite{Sommerfeld}, gives following Green's function:
\begin{equation} \label{eq:scalar_greens_function}
G(\boldsymbol r, \boldsymbol r') = \frac{e^{ik|\boldsymbol r - \boldsymbol r'|}}{4\pi |\boldsymbol r - \boldsymbol r'|}.
\end{equation}
By using the Gauss-Ostrogradski theorem, one can derive Green's theorem~\cite{Arfken},
\begin{align} \label{eq:scalar_greens_theorem}
&\iiint_V \left(U_1 \nabla^2 U_2-U_2\nabla^2 U_1\right) d^3r \notag \\
&= \oiint_S (-\bs e_n)\cdot \left(U_1\nabla U_2-U_2\nabla U_1\right)d^2r_S,
\end{align}
Where $U_1$ and $U_2$ are two scalar functions that will be filled in shortly, $V$ is a closed volume with bordering surface $S$, and vectors $\bs r_S$ lie on that surface.
Filling in $U_1=\Psi(\bs r)$ (a solution of \eqref{eq:scalar_wave_equation}) and $U_2=G(\bs r, \bs r')$ into \eqref{eq:scalar_greens_theorem}, one can obtain the \emph{Helmholtz-Kirchhoff integration theorem}:
\begin{align} \label{eq:scalar_helmholtz_kirchhoff}
\Psi(\bs r') = \frac{1}{4\pi} \oiint &\left[ \Psi(\bs r_S) \bs e_n \cdot \nabla \frac{e^{ik|\bs r_S-\bs r'|}}{|\bs r_S-\bs r'|} \right. \notag \\
&\left. - \frac{e^{ik|\bs r_S -\bs r'|}}{|\bs r_S - \bs r'|} \bs e_n \cdot \nabla \Psi(\bs r_S)  \right] d^2r_S.
\end{align}
This equation gives an exact solution to the wave equation \eqref{eq:scalar_wave_equation} in any point $\bs r'$ of $V$, given the value of $\Psi(\bs r_S)$ on the bordering surface $S$.
Being a fundamental result of a physically implied Helmholtz equation and the mathematically rigorous Gauss-Ostrogradski theorem, it finds applications in many fields of physics (albeit in a more general form).
The above can be seen as the mathematical basis of cosmologist's \emph{holographic principle}~\cite{Smoot}: information encoded on a (hyper)surface completely defines the internal state inside the entire volume enclosed by that surface.

\subsection{Boundary conditions, approximations and Fraunhofer diffraction} \label{sec:boundary_approximations}

\begin{figure}
\includegraphics[width=.75\columnwidth]{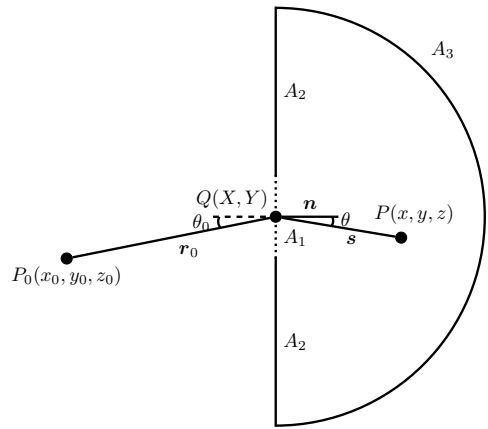}
\caption{Schematic diffraction setup. A point source in $P_0$ emits a field which illuminates an aperture $A_2$ at distance $r_0$. The wave passes through the aperture ($z=0$) at $Q$ and is detected on a screen at point $P$. The outgoing wave makes an angle $\chi$ with the incoming wave. $A_1$ is the unilluminated surrounding surface, and $A_3$ is the surface at infinity. $s$ is the distance between the aperture and focus point $P$, where the diffraction pattern is calculated. \label{fig:diffraction_setup}}
\end{figure}
The mathematical model of the diffraction setup is shown in fig. \ref{fig:diffraction_setup}.
A point source is located at $P_0$ at distance $r_0$ from an aperture $A_1$.
At point $Q\in A_1$, light can be transmitted into the right bounded volume which contains a screen at $P$ at a distance $s$ from the aperture.
We will now calculate the disturbance at point $P$ caused by the point source $P_0$.
$A_3$ is a surface that mathematically goes to infinity.
The boundary conditions used by Kirchhoff are:
\begin{itemize}
  \item $\Psi(\bs r) = \pd_{\bs r} \Psi(\bs r) = 0$ on $A_2$ and $A_3$,
  \item $\Psi(\bs r) = \frac{e^{ik \bs r}}{\bs r}$ on $A_1$, i.e. for $\bs r = \bs r_0$, which represents a field originating in a point source $P_0$ which illuminates the aperture.
\end{itemize}
Note it is the first of above conditions that breaks down mathematically; for an analytic function that has both its value and its derivate $0$ on a finite region, it must be zero everywhere, which cannot be physically true.
Rayleigh-Sommerfeld diffraction theory sets only one to zero, resulting in two solutions that are mathematically sound.
The problem is choosing the correct boundary conditions.
In reality, the Kirchhoff approximation produces very useful results nonetheless, to a high degree of accuracy in the approximations assumed here.
\\
Some additional approximations are necessary to get a useful result: we let $k r_0 \gg 1$ and $k s \gg 1$, which means that both the source and focal plane are far away from the aperture, reducing wavefront curvature effects which take more care to be included.
This leads to the \emph{Kirchhoff-Fresnel diffraction formula}:
\begin{equation} \label{eq:scalar_kirchhoff_fresnel}
\Psi(\bs r') = -\frac{ik}{4\pi} \iint_{A_1} \frac{e^{ik(r_0+s)}}{r_0 s} \frac{\cos{\theta_0}+\cos{\theta}}{2} \Psi(\bs r_{S}) dS.
\end{equation}
This is valid directly behind the aperture.
The quantity $\frac{\cos{\theta_0}+\cos{\theta}}{2}$ is called the \emph{obliquity factor}.
$\theta_0\approx 0$ is valid for a point source far away from the aperture, and $\theta\approx 0$ amounts to a small angle far field approximation.
The far field approximations and their near field improvements are discussed extensively in optics literature~\cite{Hecht}, and not of primary interest here.
For completeness, the Fraunhofer diffraction formula that one can derive by applying these approximations to \eqref{eq:scalar_kirchhoff_fresnel} can be written as follows:
\begin{equation} \label{eq:scalar_fraunhofer}
\Psi(P) \propto \iint_{A_1} e^{-i(k_x X + k_y Y)} \Psi_{\text{Ap}}(X,Y) dXdY,
\end{equation}
with $P=(k_x,k_y)$ a point in the focal plane (given in Fourier space, which can simply be rescaled to represent real coordinates in that plane), $(X,Y)$ are the coordinates on the aperture, and $\Psi_\text{Ap}$ is the wave function in the aperture.
This formula states that one can easily calculate the far-field image (including not only density but also phase) if one knows the wave function in the aperture.
Also note that this theory applies equally well to photon waves (described by the wave equation which is fully relativistically applicable) and the Schr\"odinger equation (which applies to non-relativistic matter waves).
\\
In what follows, the derivation done by P. Hillion~\cite{Hillion_Fresnel,Hillion_Helmholtz} in light of the 2-component spinor electromagnetic field is expanded to the full 4-component spinor Dirac theory for massive particles with spin.

\section{Spinor diffraction theory}

\subsection{Dirac equation} \label{sec:dirac_equation}

The Dirac equation for spin-$\frac{1}{2}$ particles is:
\begin{equation} \label{eq:dirac_equation}
\left(i \gamma^\mu \partial_\mu - m \right) \Psi = 0,
\end{equation}
which is relativistically covariant, and in most common usage a 4-component matrix equation for the Dirac spinor $\Psi$.
Applying the complex conjugate Dirac operator to the Dirac equation gives an interesting and well-known result:
\begin{align}
&\left(-i \gamma^\mu \partial_\mu - m \right)  \left(i \gamma^\mu \partial_\mu - m \right) \Psi = 0 \notag \\
\Leftrightarrow &\left(\eta^{\nu\mu} \partial_\nu \partial_\mu + m^2\right)\Psi = 0,
\end{align}
where $\eta^{\nu\mu}$ is the Minkowski metric of spacetime.
The above is nothing less than the scalar Klein-Gordon equation for every component of the Dirac spinor independently.
This property will be used in the following section.
In a stationary situation, one can write $\Psi(\bs r, t)\propto \Psi(\bs r) e^{-iEt}$, which results in
\begin{equation}
\left( \partial^j\partial_j +E^2+m^2  \right) \Psi = 0.
\end{equation}
One can now define a new variable, $k^2=m^2+E^2$, which leads to a familiar form:
\begin{equation}
\left(\partial^j \partial_j +k^2\right) \Psi = 0.
\end{equation}
The stationary Dirac equation can now be rewritten using this new variable as
\begin{equation}
\left(\gamma^j\partial_j+ik\right)\Psi=0,
\end{equation}
without loss of generality.

\subsection{Green's matrix}

Using the fact that the solutions of the Dirac equation form a subset of those of the Klein-Gordon equation, one can construct a solution to the Dirac equation $\Psi$ from scalar solutions of the Klein-Gordon equation $\psi^\mu$:
\begin{equation}
\Psi = \left(\gamma^j\partial_j -ik\right) \left(
\begin{matrix}
  \psi^0 \\ \psi^1 \\ \psi^2 \\ \psi^3
\end{matrix} \right).
\end{equation}
This is a well-known fact.
As an extension, which is also used by Hillion~\cite{Hillion_Helmholtz}, one can construct a Green's matrix solution to
\begin{equation} \label{eq:dirac_greens_equation}
-\delta(\bs r - \bs r') = g(\bs r, \bs r')\left(-i\gamma^j\partial_j +k\right),
\end{equation}
from the scalar Green's function \eqref{eq:scalar_greens_function} as follows:
\begin{equation} \label{eq:dirac_greens_matrix}
g(\bs r, \bs r') = \left(i \gamma^j\partial_j + k\right) G(\bs r, \bs r'),
\end{equation}
which is a useful form used in the next section.

\subsection{Kirchhoff diffraction}

In this section, a 4-component spinor analogue of the Helmholtz-Kirchhoff integration theorem \eqref{eq:scalar_helmholtz_kirchhoff} is derived from the Dirac equation.
Multiplying \eqref{eq:dirac_equation} with $g(\bs r, \bs r')$ on the left and subtracting that from \eqref{eq:dirac_greens_equation} multiplied on the right by $\Psi(\bs r)$, one has:
\begin{equation} \label{eq:intermediate_result}
i\partial_j \left( g(\bs r, \bs r') \gamma^j\Psi(\bs r) \right) = \Psi(\bs r) \delta(\bs r - \bs r').
\end{equation}
By using the correct mathematical formulation, one can easily prove that the Gauss-Ostrogradski theorem also applies to spinor objects~\cite{Hillion_Helmholtz,Jeffreys}, in which case one can integrate \eqref{eq:intermediate_result} over a volume $V$ around the point $\bs r'$ and write:
\begin{equation} \label{eq:intermediate_result2}
\Psi(\bs r') = -i \oiint_S g(\bs r_S,\bs r') \gamma^n \Psi(\bs r_S) d^2r_S,
\end{equation}
where the integration is over the closed surface $S$ around $V$, containing all the vectors $\bs r_S$. $\gamma^n$ is the compound $\gamma$ matrix perpendicular to the surface.
Obviously a $\gamma$ matrix cannot be perpendicular to a surface, and this is a purely notational convenience.
An example of $\gamma^n$ can easily be given in the case of a spherical surface, namely $\gamma^r=\sin{\theta}\cos{\phi}\gamma^1 + \sin{\theta}\sin{\phi}\gamma^2 +\cos{\theta}\gamma^3$, which is nothing more that the vector transformation to spherical coordinates.
It is not surprising the $\gamma$ matrices transform as vectors, as they are the basis vectors of the Dirac algebra.
This makes $\gamma^n$ more than only ``notational convenience", and a mathematical $\gamma$ object can easily be seen as orthogonal to a surface.
From \eqref{eq:scalar_greens_function} one can prove
\begin{equation}
\gamma^j\partial_j G(\bs r, \bs r') = ik \left(1+\frac{i}{ks}\right) \gamma^j G(\bs r, \bs r') \partial_j s.
\end{equation}
This expression can be used to rewrite \eqref{eq:intermediate_result2} into a form comparable to \eqref{eq:scalar_helmholtz_kirchhoff}:
\begin{align}
\Psi(\bs r') = -&\frac{ik}{4\pi} \oiint_S \frac{e^{ik|\bs r_S-\bs r'|}}{|\bs r_S - \bs r'|} \left[ 1 - \right. \notag \\
&\left. \left(1+\frac{i}{ks}\right)\gamma^j\left(\partial_j s\right) \right] \gamma^n \Psi(\bs r_S) d^2r_S.
\end{align}
The next steps are exactly the same as in the scalar theory: Kirchhoff's boundary conditions and approximations  are applied, along with:
\begin{itemize}
  \item $\gamma^r\approx \gamma^3$: when the beam is directed along the $z$ axis, $\gamma^r$ coincides with $\gamma^3$,
  \item $\partial_{r_0} s \approx -1$: valid for calculations almost directly behind the aperture.
\end{itemize}
gives the \emph{spinor Kirchhoff-Fresnel diffraction formula} (compare with \eqref{eq:scalar_kirchhoff_fresnel}):
\begin{align} \label{eq:spinor_kirchhoff_fresnel}
\Psi(\bs r')\propto \iint_{A_1} &\frac{e^{ik(r_0+s)}}{r_0 s} \left( 1+\gamma^3 \right) \gamma^3 \Psi(\bs r_{S}) dS.
\end{align}

\subsection{Fraunhofer approximation}

One can again perform the exact same approximations as in the scalar case to reduce \eqref{eq:spinor_kirchhoff_fresnel}, leading to a similar form as in \eqref{eq:scalar_fraunhofer}:
\begin{align} \label{eq:spinor_fraunhofer}
\Psi(P) \propto \iint_{A_1} &e^{-i(k_x X + k_y Y)} \left(1+\gamma^3\right)\gamma^3 \Psi_{\text{Ap}} dX dY,
\end{align}
which gives the Dirac spinor form for the far-field image of a diffracted $z$-oriented beam.

\section{Discussion}

It should not be surprising that scalar diffraction theory extends cleanly to Dirac's description of fermions.
Besides the basic properties of Dirac spinor solutions highlighted in Section \ref{sec:dirac_equation}, the 2D Pauli algebra of real space, generated by $\{\sigma^i\}$, is a subalgebra of the 4D Dirac spacetime algebra generated by $\{\gamma^\mu\}$ (both of which obey the same anticommutator relationships).
The most important properties of this result will be discussed: the non-relativistic limit and what happens to spin under diffraction.

\subsection{Spin and the non-relativistic limit}

The advantage of working in the Dirac representation is the simplicity of its non-relativistic limit.
Taking the limit $\frac{\bs k}{E+m}\rightarrow 0$, it is readily known only the top two components of the Dirac spinor contribute, and the $\gamma$-factor effectively reduces to a negligible scaling operation for a spin eigenstate.
By this, the formula \eqref{eq:spinor_fraunhofer} reduces simply to the classical form \eqref{eq:scalar_fraunhofer} in the non-relativistic case.


The theory derived here highlights the simple fact that for diffraction through a scalar aperture of a paraxial relativistic fermion beam, spin is conserved.
This is made clear by the $\gamma$ factor which is multiplied with the incoming spinor, mixing only the large and small components of the same spin elements, the explicit form for the $z$-direction given below for the Dirac representation:
\begin{equation} \label{eq:gamma_factor}
\left(
\begin{matrix}
-1 & 0 & 1 & 0 \\
0 & -1 & 0 & -1 \\
-1 & 0 & -1 & 0 \\
0 & 1 & 0 & -1
\end{matrix}\right).
\end{equation}
For the other directions, the nonzero elements of the $\gamma$-factor align with the nonzero elements of the plane wave spinor with its spin aligned in that direction.
For example, the spinor part of a plane wave eigenstate of $\Sigma_z=+1$ is given by
\begin{equation}
\left(
\begin{matrix}
1 \\ 0 \\
\frac{k_z}{E+m} \\ 0
\end{matrix}\right),
\end{equation}
which, after multiplication with $(1+\gamma^3)\gamma^3$, is still in the same eigenstate of $\Sigma_z$.
The Fourier transform will not affect the spinor structure.
Hence, the far field spinor given by \eqref{eq:spinor_fraunhofer} is also unmodified with respect to spin after diffraction.
This result implies that a pure diffraction experiment with a scalar aperture will not be influenced by electron spin, at least in the paraxial region.
Some mixing may occur far from the beam axis, where $\left(1+\gamma^r\right)\gamma^r$ is not well approximated by (for the $z$ direction) $\left(1+\gamma^3\right)\gamma^3$.
The generality of the results derived here allow to extend the simple approximation \eqref{eq:spinor_fraunhofer} to more complicated scenarios, including any type of spin polarization.

\subsection{Kirchhoff and Rayleigh-Sommerfeld boundary conditions}

It is well known that Kirchhoff's boundary conditions lead to physical solutions containing a grave mathematical inconsistency.
By putting both the wave function and its derivative to zero on a finite region of space, one has inevitably that $\Psi$, an analytic function, must be zero everywhere inside the bounding surface (this includes all points $P$).
Rayleigh and Sommerfeld solved this by deriving two possible integration formulas, each for different boundary conditions: one has only the wavefunction zero, the other has only its derivative zero on the ``dark areas".
Table \ref{tab:kirchhoff_vs_rayleighsommerfeld} attempts to give an overview of the conceptual differences between the three integral formalisms.
\begin{table}
  \begin{tabular}{c | c | c}
    Kirchhoff & R-S 1 & R-S 2 \\
    \hline
    $\pd_{\bs r} \Psi(\bs r) = \Psi(\bs r) = 0$ & $\pd_{\bs r} \Psi(\bs r) = 0$ & $\Psi(\bs r) = 0$ \\
    \hline
    $e^{ikr}/r$ & $e^{ikr}/r-e^{iks}/s$ & $e^{ikr}/r+e^{iks}/s$ \\
    \hline
    $(\cos{\theta_0} + \cos{\theta})/2$ & $\cos{\theta}$ & $\cos{\theta_0}$
  \end{tabular}
  \caption{Comparison between essential differences between scalar Kirchhoff and Rayleigh-Somerfeld theories. The first row shows the boundary conditions applied on $A_{1,3}$ in Figure \ref{fig:diffraction_setup}. The second row shows the Green's function used. The third row show the resulting obliquity factor.\label{tab:kirchhoff_vs_rayleighsommerfeld}}
\end{table}
The \emph{Kirchhoff-Helmholtz integration theorems} for these cases under the present approximations differ slightly:
\begin{subequations}
  \begin{align}
  \text{Kirchoff: } &\frac{1}{4\pi} \oiint_S \left( (\pd_{\bs r} \Psi) G - \Psi \pd_{\bs r} G\right) d\bs S \\
  \text{R-S 1: } &-\frac{1}{2\pi} \oiint_S \Psi \pd_{\bs r} G d\bs S \\
  \text{R-S 2: } &\frac{1}{2\pi} \oiint_S (\pd_{\bs r} \Psi) G d\bs S
  \end{align}
\end{subequations}
In the Dirac case, the Green's functions become the Green's matrices derived from the scalar Green's functions, through \eqref{eq:dirac_greens_matrix}.
Note that one can also apply the conjugate Dirac operator in that equation, and the resulting Green's matrix will be antisymmetric to the one used here (see~\cite{Hillion_Helmholtz}), which is expected from the anticommutator relationship fullfilled by the Dirac spinors.

\section{Conclusion}

The formal scalar Kirchhoff diffraction theory is shown to extend mathematically to relativistic Dirac spinors, starting from the Dirac equation.
The entire derivation is shown, and the result is consistent with the scalar theory.
The mathematical fallacy present in the Kirchhoff derivation is explained, and the effect on the results derived here is elucidated.
The non-relativistic limit is found to reduce to the classical results.
The effect of spin is analyzed by using the spinor Fraunhofer diffraction equation, where it is evident that straight propagation through an aperture does not alter the spin eigenstate.
More oblique incidence and the wave solution at a larger angle from the straight line propagation considered here will show differences, governed by the respective directional $\gamma$ matrix and the incoming spinor wave.



%
%

%

\begin{acknowledgments}
Acknowledgements:
\\
This research was supported by a PhD fellowship grant from the FWO (Aspirant Fonds Wetenschappelijk Onderzoek - Vlaanderen).
The authors acknowledge financial support from the European Union under the Seventh Framework Program under a contract for an Integrated Infrastructure Initiative.
Reference No. 312483-ESTEEM2. J. Verbeeck acknowledges funding from the European Research Council under the 7th Framework Program (FP7), ERC grant N246791-COUNTATOMS and ERC Starting Grant 278510 VORTEX.
\end{acknowledgments}

\bibliography{bibliography}

\begin{thebibliography}{18}%
\makeatletter
\providecommand \@ifxundefined [1]{%
 \@ifx{#1\undefined}
}%
\providecommand \@ifnum [1]{%
 \ifnum #1\expandafter \@firstoftwo
 \else \expandafter \@secondoftwo
 \fi
}%
\providecommand \@ifx [1]{%
 \ifx #1\expandafter \@firstoftwo
 \else \expandafter \@secondoftwo
 \fi
}%
\providecommand \natexlab [1]{#1}%
\providecommand \enquote  [1]{``#1''}%
\providecommand \bibnamefont  [1]{#1}%
\providecommand \bibfnamefont [1]{#1}%
\providecommand \citenamefont [1]{#1}%
\providecommand \href@noop [0]{\@secondoftwo}%
\providecommand \href [0]{\begingroup \@sanitize@url \@href}%
\providecommand \@href[1]{\@@startlink{#1}\@@href}%
\providecommand \@@href[1]{\endgroup#1\@@endlink}%
\providecommand \@sanitize@url [0]{\catcode `\\12\catcode `\$12\catcode
  `\&12\catcode `\#12\catcode `\^12\catcode `\_12\catcode `\%12\relax}%
\providecommand \@@startlink[1]{}%
\providecommand \@@endlink[0]{}%
\providecommand \url  [0]{\begingroup\@sanitize@url \@url }%
\providecommand \@url [1]{\endgroup\@href {#1}{\urlprefix }}%
\providecommand \urlprefix  [0]{URL }%
\providecommand \Eprint [0]{\href }%
\providecommand \doibase [0]{http://dx.doi.org/}%
\providecommand \selectlanguage [0]{\@gobble}%
\providecommand \bibinfo  [0]{\@secondoftwo}%
\providecommand \bibfield  [0]{\@secondoftwo}%
\providecommand \translation [1]{[#1]}%
\providecommand \BibitemOpen [0]{}%
\providecommand \bibitemStop [0]{}%
\providecommand \bibitemNoStop [0]{.\EOS\space}%
\providecommand \EOS [0]{\spacefactor3000\relax}%
\providecommand \BibitemShut  [1]{\csname bibitem#1\endcsname}%
\let\auto@bib@innerbib\@empty
\bibitem [{\citenamefont {H\'enault}(2010)}]{Henault}%
  \BibitemOpen
  \bibfield  {author} {\bibinfo {author} {\bibfnamefont {F.}~\bibnamefont
  {H\'enault}},\ }\href {\doibase 10.1364/JOSAA.27.000435} {\bibfield
  {journal} {\bibinfo  {journal} {J. Opt. Soc. Am. A}\ }\textbf {\bibinfo
  {volume} {27}},\ \bibinfo {pages} {435} (\bibinfo {year} {2010})}\BibitemShut
  {NoStop}%
\bibitem [{\citenamefont {Takeda}\ \emph {et~al.}(1982)\citenamefont {Takeda},
  \citenamefont {Ina},\ and\ \citenamefont {Kobayashi}}]{Takeda}%
  \BibitemOpen
  \bibfield  {author} {\bibinfo {author} {\bibfnamefont {M.}~\bibnamefont
  {Takeda}}, \bibinfo {author} {\bibfnamefont {H.}~\bibnamefont {Ina}}, \ and\
  \bibinfo {author} {\bibfnamefont {S.}~\bibnamefont {Kobayashi}},\ }\href
  {\doibase 10.1364/JOSA.72.000156} {\bibfield  {journal} {\bibinfo  {journal}
  {J. Opt. Soc. Am.}\ }\textbf {\bibinfo {volume} {72}},\ \bibinfo {pages}
  {156} (\bibinfo {year} {1982})}\BibitemShut {NoStop}%
\bibitem [{\citenamefont {Takeda}\ and\ \citenamefont {Mutoh}(1983)}]{Takeda2}%
  \BibitemOpen
  \bibfield  {author} {\bibinfo {author} {\bibfnamefont {M.}~\bibnamefont
  {Takeda}}\ and\ \bibinfo {author} {\bibfnamefont {K.}~\bibnamefont {Mutoh}},\
  }\href {\doibase 10.1364/AO.22.003977} {\bibfield  {journal} {\bibinfo
  {journal} {Appl. Opt.}\ }\textbf {\bibinfo {volume} {22}},\ \bibinfo {pages}
  {3977} (\bibinfo {year} {1983})}\BibitemShut {NoStop}%
\bibitem [{\citenamefont {Verbeeck}\ \emph {et~al.}(2010)\citenamefont
  {Verbeeck}, \citenamefont {Tian},\ and\ \citenamefont
  {Schattschneider}}]{Verbeeck}%
  \BibitemOpen
  \bibfield  {author} {\bibinfo {author} {\bibfnamefont {J.}~\bibnamefont
  {Verbeeck}}, \bibinfo {author} {\bibfnamefont {H.}~\bibnamefont {Tian}}, \
  and\ \bibinfo {author} {\bibfnamefont {P.}~\bibnamefont {Schattschneider}},\
  }\href {\doibase 10.1038/nature09366} {\bibfield  {journal} {\bibinfo
  {journal} {Nature}\ }\textbf {\bibinfo {volume} {467}},\ \bibinfo {pages}
  {301} (\bibinfo {year} {2010})}\BibitemShut {NoStop}%
\bibitem [{\citenamefont {Verbeeck}\ \emph {et~al.}(2012)\citenamefont
  {Verbeeck}, \citenamefont {Tian},\ and\ \citenamefont
  {B\'ech\'e}}]{Verbeeck_spiral}%
  \BibitemOpen
  \bibfield  {author} {\bibinfo {author} {\bibfnamefont {J.}~\bibnamefont
  {Verbeeck}}, \bibinfo {author} {\bibfnamefont {H.}~\bibnamefont {Tian}}, \
  and\ \bibinfo {author} {\bibfnamefont {A.}~\bibnamefont {B\'ech\'e}},\ }\href
  {\doibase 10.1016/j.ultramic.2011.10.008} {\bibfield  {journal} {\bibinfo
  {journal} {Ultramicroscopy}\ }\textbf {\bibinfo {volume} {113}},\ \bibinfo
  {pages} {83} (\bibinfo {year} {2012})}\BibitemShut {NoStop}%
\bibitem [{\citenamefont {Schattschneider}\ \emph {et~al.}(2012)\citenamefont
  {Schattschneider}, \citenamefont {St\"ger-Pollach}, \citenamefont
  {L\"offler}, \citenamefont {Steiger-Thirsfeld}, \citenamefont {Hell},\ and\
  \citenamefont {Verbeeck}}]{Schattschneider_fork}%
  \BibitemOpen
  \bibfield  {author} {\bibinfo {author} {\bibfnamefont {P.}~\bibnamefont
  {Schattschneider}}, \bibinfo {author} {\bibfnamefont {M.}~\bibnamefont
  {St\"ger-Pollach}}, \bibinfo {author} {\bibfnamefont {S.}~\bibnamefont
  {L\"offler}}, \bibinfo {author} {\bibfnamefont {A.}~\bibnamefont
  {Steiger-Thirsfeld}}, \bibinfo {author} {\bibfnamefont {J.}~\bibnamefont
  {Hell}}, \ and\ \bibinfo {author} {\bibfnamefont {J.}~\bibnamefont
  {Verbeeck}},\ }\href {\doibase 10.1016/j.ultramic.2012.01.010} {\bibfield
  {journal} {\bibinfo  {journal} {Ultramicroscopy}\ }\textbf {\bibinfo {volume}
  {115}},\ \bibinfo {pages} {21} (\bibinfo {year} {2012})}\BibitemShut
  {NoStop}%
\bibitem [{\citenamefont {Bliokh}\ \emph {et~al.}(2007)\citenamefont {Bliokh},
  \citenamefont {Bliokh}, \citenamefont {Savel'ev},\ and\ \citenamefont
  {Nori}}]{Bliokh_semiclassical}%
  \BibitemOpen
  \bibfield  {author} {\bibinfo {author} {\bibfnamefont {K.~Y.}\ \bibnamefont
  {Bliokh}}, \bibinfo {author} {\bibfnamefont {Y.~P.}\ \bibnamefont {Bliokh}},
  \bibinfo {author} {\bibfnamefont {S.}~\bibnamefont {Savel'ev}}, \ and\
  \bibinfo {author} {\bibfnamefont {F.}~\bibnamefont {Nori}},\ }\href {\doibase
  10.1103/PhysRevLett.99.190404} {\bibfield  {journal} {\bibinfo  {journal}
  {Phys. Rev. Lett.}\ }\textbf {\bibinfo {volume} {99}},\ \bibinfo {pages}
  {190404} (\bibinfo {year} {2007})}\BibitemShut {NoStop}%
\bibitem [{\citenamefont {Schattschneider}\ and\ \citenamefont
  {Verbeeck}(2011)}]{Schattschneider_theory}%
  \BibitemOpen
  \bibfield  {author} {\bibinfo {author} {\bibfnamefont {P.}~\bibnamefont
  {Schattschneider}}\ and\ \bibinfo {author} {\bibfnamefont {J.}~\bibnamefont
  {Verbeeck}},\ }\href {\doibase 10.1016/j.ultramic.2011.07.004} {\bibfield
  {journal} {\bibinfo  {journal} {Ultramicroscopy}\ }\textbf {\bibinfo {volume}
  {111}},\ \bibinfo {pages} {1461} (\bibinfo {year} {2011})}\BibitemShut
  {NoStop}%
\bibitem [{\citenamefont {Janicijevic}\ and\ \citenamefont
  {Topuzoski}(2008)}]{Janicijevic}%
  \BibitemOpen
  \bibfield  {author} {\bibinfo {author} {\bibfnamefont {L.}~\bibnamefont
  {Janicijevic}}\ and\ \bibinfo {author} {\bibfnamefont {S.}~\bibnamefont
  {Topuzoski}},\ }\href {\doibase 10.1364/JOSAA.25.002659} {\bibfield
  {journal} {\bibinfo  {journal} {J. Opt. Soc. Am. A}\ }\textbf {\bibinfo
  {volume} {25}},\ \bibinfo {pages} {2659} (\bibinfo {year}
  {2008})}\BibitemShut {NoStop}%
\bibitem [{\citenamefont {Hillion}\ and\ \citenamefont
  {Quinnez}(1983)}]{Hillion_Fresnel}%
  \BibitemOpen
  \bibfield  {author} {\bibinfo {author} {\bibfnamefont {P.}~\bibnamefont
  {Hillion}}\ and\ \bibinfo {author} {\bibfnamefont {S.}~\bibnamefont
  {Quinnez}},\ }\href {\doibase 10.1088/0150-536X/14/3/002} {\bibfield
  {journal} {\bibinfo  {journal} {Journal of Optics}\ }\textbf {\bibinfo
  {volume} {14}},\ \bibinfo {pages} {143} (\bibinfo {year} {1983})}\BibitemShut
  {NoStop}%
\bibitem [{\citenamefont {Hillion}(1978)}]{Hillion_Helmholtz}%
  \BibitemOpen
  \bibfield  {author} {\bibinfo {author} {\bibfnamefont {P.}~\bibnamefont
  {Hillion}},\ }\href {\doibase 10.1063/1.523547} {\bibfield  {journal}
  {\bibinfo  {journal} {Journal of Mathematical Physics}\ }\textbf {\bibinfo
  {volume} {19}},\ \bibinfo {pages} {264} (\bibinfo {year} {1978})}\BibitemShut
  {NoStop}%
\bibitem [{\citenamefont {Hillion}(1979)}]{Hillion_geometric}%
  \BibitemOpen
  \bibfield  {author} {\bibinfo {author} {\bibfnamefont {P.}~\bibnamefont
  {Hillion}},\ }\href {http://stacks.iop.org/0150-536X/10/i=1/a=003} {\bibfield
   {journal} {\bibinfo  {journal} {Journal of Optics}\ }\textbf {\bibinfo
  {volume} {10}},\ \bibinfo {pages} {21} (\bibinfo {year} {1979})}\BibitemShut
  {NoStop}%
\bibitem [{\citenamefont {Hillion}\ and\ \citenamefont
  {Quinnez}(1987)}]{Hillion_spinor}%
  \BibitemOpen
  \bibfield  {author} {\bibinfo {author} {\bibfnamefont {P.}~\bibnamefont
  {Hillion}}\ and\ \bibinfo {author} {\bibfnamefont {S.}~\bibnamefont
  {Quinnez}},\ }\href {http://stacks.iop.org/0150-536X/18/i=2/a=001} {\bibfield
   {journal} {\bibinfo  {journal} {Journal of Optics}\ }\textbf {\bibinfo
  {volume} {18}},\ \bibinfo {pages} {51} (\bibinfo {year} {1987})}\BibitemShut
  {NoStop}%
\bibitem [{\citenamefont {Sommerfeld}(1949)}]{Sommerfeld}%
  \BibitemOpen
  \bibfield  {author} {\bibinfo {author} {\bibfnamefont {A.}~\bibnamefont
  {Sommerfeld}},\ }\href@noop {} {\emph {\bibinfo {title} {Partial Differential
  Equations in Physics}}}\ (\bibinfo  {publisher} {Academic Press},\ \bibinfo
  {year} {1949})\BibitemShut {NoStop}%
\bibitem [{\citenamefont {Arfken}\ \emph {et~al.}(2005)\citenamefont {Arfken},
  \citenamefont {Weber},\ and\ \citenamefont {Harris}}]{Arfken}%
  \BibitemOpen
  \bibfield  {author} {\bibinfo {author} {\bibfnamefont {G.~B.}\ \bibnamefont
  {Arfken}}, \bibinfo {author} {\bibfnamefont {H.~J.}\ \bibnamefont {Weber}}, \
  and\ \bibinfo {author} {\bibfnamefont {F.~E.}\ \bibnamefont {Harris}},\
  }\href@noop {} {\emph {\bibinfo {title} {Mathematical Methods for Physicists,
  6th ed.}}}\ (\bibinfo  {publisher} {Elsevier Academic Press},\ \bibinfo
  {year} {2005})\ pp.\ \bibinfo {pages} {60--63}\BibitemShut {NoStop}%
\bibitem [{\citenamefont {Smoot}(2010)}]{Smoot}%
  \BibitemOpen
  \bibfield  {author} {\bibinfo {author} {\bibfnamefont {G.~F.}\ \bibnamefont
  {Smoot}},\ }\href {\doibase 10.1142/S0218271810018414} {\bibfield  {journal}
  {\bibinfo  {journal} {International Journal of Modern Physics D}\ }\textbf
  {\bibinfo {volume} {19}},\ \bibinfo {pages} {2247} (\bibinfo {year}
  {2010})}\BibitemShut {NoStop}%
\bibitem [{\citenamefont {Hecht}(2002)}]{Hecht}%
  \BibitemOpen
  \bibfield  {author} {\bibinfo {author} {\bibfnamefont {E.}~\bibnamefont
  {Hecht}},\ }\href@noop {} {\emph {\bibinfo {title} {Optics (Fourth
  Edition)}}}\ (\bibinfo  {publisher} {Addison Wesley},\ \bibinfo {year}
  {2002})\BibitemShut {NoStop}%
\bibitem [{\citenamefont {Jeffreys}\ and\ \citenamefont
  {Swirles}(1972)}]{Jeffreys}%
  \BibitemOpen
  \bibfield  {author} {\bibinfo {author} {\bibfnamefont {H.}~\bibnamefont
  {Jeffreys}}\ and\ \bibinfo {author} {\bibfnamefont {B.}~\bibnamefont
  {Swirles}},\ }\href@noop {} {\emph {\bibinfo {title} {Methods of Mathematical
  Physics, third edition}}}\ (\bibinfo  {publisher} {University Press},\
  \bibinfo {year} {1972})\BibitemShut {NoStop}%
\end{thebibliography}%

\end{document}